\documentclass[pre,aps,twocolumn,superscriptaddress,showpacs,floatfix,amsmath,amssym]{revtex4}
\usepackage{graphicx}
\usepackage{subfigure}
\usepackage{amssymb}
\usepackage{epstopdf}

\newcommand{\cH}{{\cal H}}

\newcommand{\cA}{{\cal A}}

\newcommand{\cV}{{\cal V}}
\newcommand{\ep}{\epsilon}
\newcommand{\al}{\alpha}

\begin{document}

\title{Undulated cylinders of charged diblock copolymers}

\author{Gregory M. Grason}
\affiliation{Department of Physics and Astronomy, University of Pennsylvania, Philadelphia, PA  19104, USA}
\affiliation{Department of Physics and Astronomy, University of California, Los Angeles, CA  90095, USA}
\author{Christian D. Santangelo}
\email{santancd@physics.upenn.edu}
\affiliation{Department of Physics and Astronomy, University of Pennsylvania, Philadelphia, PA  19104, USA}
\date{\today}

\begin{abstract}
We study the cylinder to sphere morphological transition of diblock copolymers in aqueous solution with a hydrophobic block and a charged block.  We find a metastable undulated cylinder configuration for a range of charge and salt concentrations which, nevertheless, occurs above the threshold where spheres are thermodynamically favorable.  By modeling the shape of the cylinder ends, we find that the free energy barrier for the transition from cylinders to spheres is quite large and that this barrier falls significantly in the limit of high polymer charge and low solution salinity.  This suggests that observed undulated cylinder phases are kinetically trapped structures.
\end{abstract}


\maketitle

Diblock copolymers, formed by joining two immiscible polymers at one end, are a model system to study many aspects of self-assembly and order-order transitions from one self-assembled state to another~\cite{bates_sci_91, bates_fred_arpc_90, fredrickson}.  Self-assembly in diblock copolymers is driven by two competing effects: the immiscibility of the two blocks drives their phase separation, yet the connectivity of the blocks prevents a true macroscopic phase separation.  Even in the simplest case of copolymer melts, this competition leads to the formation of a variety of generic structures, including spherical and cylindrical micelles, lamellae and bicontinuous gyroid phase~\cite{matsen_schick_prl_94, semenov_jetp_85}.  The selection of these various morphologies is determined by a balance of surface tension between microphase separated regions and the chain stretching of the polymers and can, in certain cases, be analyzed purely from the point of view of the interfacial geometry~\cite{grason}.

In selective solvents, on the other hand,  the morphology of copolymer structures is also affected by solvent-polymer interactions, widening the both the range of possible structures and the ability to control morphological transitions by tuning the solvent conditions.   Moreover, in solution, the volume-filling constraints of the molten state are relaxed.  In this way, diblock copolymers are the analogues of surfactants, and exhibit many of the typical self-assembled morphologies: spherical micelles, wormlike (cylindrical) micelles, and lamellae~\cite{israelachvili}.  Despite the similarities, however, these systems also display a variety of novel structures, including wormlike micelle networks~\cite{jain_bates_sci_03, jain_bates_macro_04, eisenberg1}, large compound vesicles~\cite{discher_eisenberg_sci_02, eisenberg1, discher_jpcb_05}, and undulated cylinders~\cite{jain_bates_sci_03,jain_bates_macro_04, eisenberg1, bendejacq}.

In this paper, we study the undulated cylindrical morphology by considering axisymmetric, diblock copolymer micelles dispersed in aqueous solution, concentrating on the case of copolymers composed of a hydrophobic block and a polyelectrolyte block.  In part, this focus is due to the fact the self-assembly in polyelectrolyte copolymer solutions has been well characterized experimentally \cite{discher_jpcb_05}.  While similar experiments have address the morphological transition between wormlike micelles and spheres for neutral block copolymers~\cite{eisenberg2, jain_bates_macro_04}, the added control provided by modulating the charge and salt conditions of the polyelectrolyte domains allow for a systematic investigation of, in particular, the cylinder to sphere (CS) transition.   Bendejacq {\it et al.}~\cite{bendejacq} have performed such experiments at controlled salinity and ionic strength and find that when the polyelectrolyte domains are highly charged, the CS transition proceeds by the undulation of the cylindrical micelles followed by the eventual breakup into spheres.  At certain intermediate ionic strengths and salt concentrations, however, undulated cylinders were also observed to have very long lifetimes, and were suggested as a candidate for a new intermediate phase between cylinders and spheres in PS-PAA diblock copolymers.  

We develop a mean-field model for diblock copolymer micelles with interfacial geometry between straight cylinders and individual spheres.  Experimentally, this can be realized by first forming the cylinders, then changing solvent conditions to frustrate the geometry toward the formation of spherical micelles, as was the method in the experiments of ref. \cite{bendejacq}.  To this end, we construct a class of micellar morphologies whose geometry can be continuously varied from cylindrical to spherical and compute the free energy within a variational, mean-field scheme.  Our main result is that under certain salt and charge conditions, cylindrical micelles become unstable to the formation of undulated cylinders.  These undulated cylinders are metastable, however, and their observation by Bendejacq \textit{et al.}~\cite{bendejacq} must indicate a large kinetic barrier to the breakup of undulated cylinders to spheres.  Our model allows us to calculate an upper bound on the free-energy barrier for fission of a cylinder into spheres, and we find a strong dependence of the barrier height on salt concentration of the solution and the effective charge of the polyelectrolyte blocks.  Combined these calculations enable us qualitatively to understand the ``phase diagram'' observed in reference~\cite{bendejacq}.  We also discuss results on the shape of the micelle ends.

Our approach is to construct the simplest possible model for the energetics of undulated cylinders that contains the essential physics.  In this spirit, we will ignore the fluctuations of the chains around their primitive paths and assume the chains in the corona are equally stretched.  Strictly speaking, this analysis provides only an upper-bound on the true free energy.  Nevertheless, similar calculations are known to be very useful in understanding the equilibrium morphologies of diblock copolymer melts \cite{olmsted_milner}.  Moreover, for \textit{charged} polymers in particular, which tend to be strongly-stretched, these techniques have been developed extensively by Borisov, Zhulina and collaborators~\cite{borisov-zhulina, netz}.  Recently, Jusufi \textit{et al.}~\cite{jusufi} have performed extensive simulations and detailed analytical analysis on charged star polymers and found excellent agreement between simulations and a mean-field electrostatic theory.

\begin{figure}[b]
\begin{center}
\resizebox{3in}{!}{\includegraphics{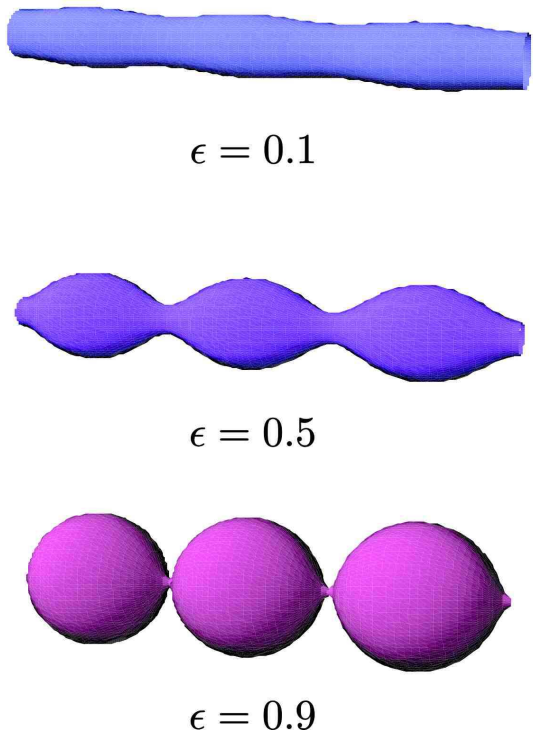}}
\caption{The unduloid family of surfaces for some particular values of $\epsilon$, spanning shapes from cylinders to a necklace of spheres.  These surfaces have constant mean curvature given by $(2 R_c)^{-1}$, where $R_c$ is  the radius of the cylindrical state at  $\epsilon=0$.}
\label{fig:unduloid}
\end{center}
\end{figure}

The paper is organized as follows: In section~\ref{sec:model} we discuss the details of our model for the diblock copolymers.  In section~\ref{sec:results}, we present phase diagrams and results on a metastable undulated cylinder phase.  We also discuss the shape of the cylinder ends and compute estimates for the free energy barriers.  In section~\ref{sec:discussion}, we discuss our results.

\section{Model for undulated cylinders}
\label{sec:model}
As a model for the interface we consider the unduloid, a family of constant mean curvature surfaces that appear in the study of soap films~\cite{kenmotsu}.  These have also been used to describe the pearling instability in fluid membrane tubes~\cite{mladenov_epjb_02}.  For our purposes, the unduloid surfaces have the convenient property that their geometry can be tuned continuously and rather simply from cylindrical to spherical.  This family of surfaces is described by two parameters: $\epsilon$ which varies from $0$ for the cylinder to $1$ for a necklace of spheres, and the size scale $R_c$ which sets the radius of the cylinder at $\epsilon=1$ \cite{kenmotsu}.    Intermediate values of $\epsilon$ describe an undulated cylinder surface (see Fig.~\ref{fig:unduloid}).   Following ref. \cite{mladenov_epjb_02}, the surface can be represented parametrically with a vector that maps $u$, the coordinate along the cylinder axis, and $\phi$, the azimuthal coordinate, to points on the interface between the charged and hydrophobic blocks, $\textbf{X}(u, \phi) = z(u) \hat{z} + r(u) \hat{r}$, where
\begin{equation}
\label{eq: z}
z(u) = R_c \int_{-\pi/2}^u dt~\frac{1+\epsilon \sin t}{\sqrt{1 + 2 \epsilon \sin t + \epsilon^2}},
\end{equation}
and
\begin{equation}
\label{eq: r}
r(u) = R_c \sqrt{1+2 \epsilon \sin u + \epsilon^2}.
\end{equation}
We see that ${\bf X}(u, \phi)$ is periodic as $u \rightarrow u+ 2 \pi$.  It is important to note that, while $R_c$ describes the radius of the cylinders, the radii of the spheres at $\epsilon=1$ is $2 R_c$.   Note that at small $\ep$, the unduloid describes a simple sinusoidal modulation of the cylinder radius,
\begin{equation}
r(z)/R_c = 1+ \ep \cos (1/R_c) + O(\ep^2) \ .
\end{equation}
In the case of soap films, unduloids are necessarily unstable due to their finite length.  For diblock copolymers, however, we will see the situation is far more complicated.

The free energy of polyelectroyte copolymer micelles is a sum of four terms,
\begin{equation}
F = F_{c,st} + F_{int} + F_{b,st} + F_{ci},
\end{equation}
where $F_{c,st}$ is the stretching energy of the chains in the core, $F_{int}$ is the interfacial energy between the core and dissolved brush domains, $F_{b,st}$ is the stretching energy of the brush, and $F_{ci}$ is the contribution to the free energy from the charged groups on the chain and the ions in solution.  In the remainder of this section we provide results and motivation for the expressions, leaving their detailed evaluation to the appendix.

\begin{figure*}
\begin{center}
\resizebox{6.5in}{!}{\includegraphics{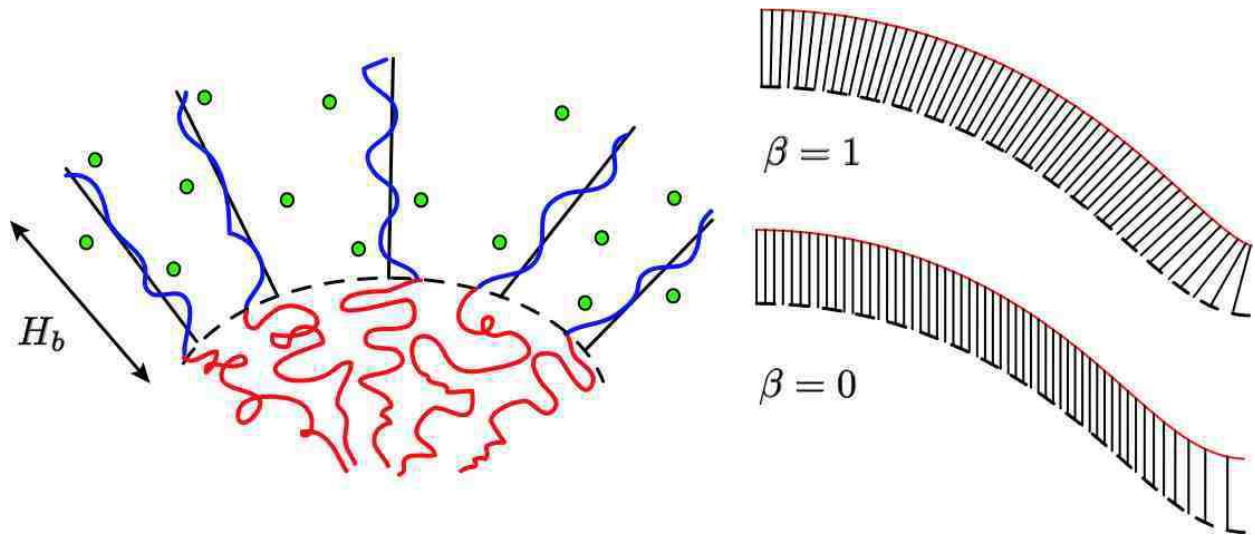}}
\caption{On the left, a schematic of the structure of the strongly-stretched corona.  The corona chains are approximated by the equal stretching hypothesis of Gaussian chains along the average trajectories (dashed).  On the right, the chains in the corona are normal to the interface surface for $\beta = 1$ and radial for $\beta=0$.  The density of chains is indicated by the density of primitive paths.  The chain density is least in the undulated necks because fewest chains reach the interface there from the core region.}
\label{fig:model}
\end{center}
\end{figure*}

\subsection{Micelle Core and Interface}
If the polymer block composing the core region is sufficiently hydrophobic, then the concentration of solvent in the core region is negligible and the chains in the core are in the molten state with constant density $\rho_0$.  Thus, the techniques of self-consistent strong-segregation theory of diblock copolymers in a melt can be applied \cite{semenov_jetp_85}.  In the strong-segregation limit, the width interface between domains is small compared to the size of the domains, and the free energy can be expressed in terms of the geometry of the interface.  To this end we employ the ``wedge model'' of Milner and Olmsted~\cite{olmsted_milner} in which we partition the core region--contained by the unduloid surface--into a series of wedges small enough so that all chains within effectively experience a single, fixed geometry.  The total stretching free energy  is computed by,
\begin{equation}
\frac{F_{c,st}}{k_B T} = \frac{3 \pi^2 \rho_0}{8 N_c^2 a^2} \int_{V_c} dV~h_c^2,
\end{equation}
where $\int_{V_c} dV$ denotes an integral over the volume of the core domain, appropriately apportioned into infinitesimal wedges, $h_c$ is the distance along the chain trajectory to the interface, $\rho_0$ is the density of monomers in the core, $N_c$ is the molecular weight of the chains in the core and $a$ is the Kuhn length.  We assume that the chains meet the interface along the surface normal.  One advantage of using the unduloid representation of the interface rather than a cylinder with a cosine modulation of the radius is that, for all values of $\epsilon$, the chain paths do not cross, allowing us to follow the transition all the way from the cylinder morphology at $\epsilon=0$ to a chain of spheres at $\epsilon=1$.

We show in the appendix that the stretching energy per chain can be put into the form
\begin{equation}
\label{eq: cst}
\frac{F_{c,st}}{n k_B T} = \frac{\pi R_c^2}{8 N_c a^2} \frac{\mathcal{S}(\epsilon)}{\cV (\epsilon)},
\end{equation}
where $n$ is the number of chains, $\cV(\epsilon) = V/\pi^2 R_c^3$ is the dimensionless volume of the micelle core, $V$ is the core volume, and $\mathcal{S}(\epsilon)$ is the dimensionless stretching energy for the chains in the core.  Since this term represents the entropic cost of confining a chain within a brush, the free energy simply scales as $R_c^2/N_c a^2$, where $N_c a^2$ is the unperturbed radius of gyration of the polymer block.

At the interface between the core and dissolved brush domains, the hydrophobic block is exposed to solvent, as well as the polyelectrolyte block.  Assuming that the interface is sufficiently narrow, we can capture energetic cost of this exposure by introducing a free energy per unit area, or surface tension, $\gamma$.  In general, such a surface tension will have a complicated dependence on the relative concentrations of solvent and polymer at the interface, free charge and salt concentration, {\it etc.}  We will treat, $\gamma$, as an unknown phenomenological parameter effectively fixed for a given system at constant temperature.  The interfacial energy per chain is then given by
\begin{equation}
\label{eq: int}
\frac{F_{int}}{n k_B T} = \frac{\gamma A(\epsilon, R_c)}{n} = \frac{2 \gamma N_c \cA (\epsilon)}{\rho_0 R_c \cV(\epsilon)},
\end{equation}
where $\gamma$ is the surface tension (in units of $k_B T$), $A(\ep,R_c)$ is the interface area, and $\cA(\epsilon) = A (\ep, R_c) /\pi^2 R_c^2$ the dimensionless area of the unduloid surface.  

\subsection{Micelle Corona}
In the limit that the concentration of free ions (both counter ions and salt ions) in the brush is large, the primary effect of these free charges is to maintain electroneutrality \cite{borisov-zhulina}.  In this limit, the so-called ``osmotic-brush" regime, the stretching of the chains is balanced by the entropy of the counterions confined to the brush, and the counterions locally neutralize the charge on the chains \cite{fyl}.  In the present calculation, we consider chains that are equally-stretched~\cite{alexander-deGennes}, and neglect polymer trajectory fluctuations.  The latter approximation can be justified in the low-salt limit and the limit of large polyelectrolyte block monomer number, $N_b$.  Fluctuations of chain path around the mean trajectory have characteristic size, $N_b^{1/2} a$, while simple scaling considerations show that in the low-salt limit the brush height grows as $H_b\sim N_b a$ \cite{netz}.  Thus, in the asymptotic, large $N_b$ limit, such fluctuations can be ignored.  Further, Jusufi \textit{et al.}~\cite{jusufi} have found that these these approximations lead to reasonable quantitative agreement with simulations for star polymers because the osmotic pressure of the counterions lead to highly stretched chains.  However, their model includes two effects which we neglect here: (1) the condensation of counterions, which we simply absorb into a redefinition of the fraction of charged monomers, $\alpha$, and (2) the impartial neutralization of the brush due to the escape of counterions to solution.

A schematic of the chain model is shown is figure~\ref{fig:model}.  We assume a brush height, $H_b$, and Gaussian chains with stretching energy
\begin{equation}
\frac{F_{b,st}}{k_B T} = \frac{3}{2 N_b a^2} \int dA~H_b^2,
\end{equation}
where for simplicity we take the Kuhn length of the polyelectrolyte block to be, $a$, as it is for the insoluble block.  

To allow for appropriate generality, we introduce into our {\it ansatz} another parameter $\beta$  which describes the average direction of the chains in the brush,
\begin{equation}
\hat{D}_b = \frac{\beta \hat{n} + (1-\beta) \hat{r}}{\sqrt{\beta^2 + (1-\beta)^2 + 2 \beta (1-\beta) \hat{n} \cdot \hat{r}}},
\end{equation}
where $\hat{n}$ is the unit normal of the interface (see Fig. \ref{fig:model}).  In the regions of high negative mean curvature, the density of chains, and hence the osmotic pressure, is much larger than the rest of the brush.  Since charged brushes do not typically overlap~\cite{brushoverlap}, we  can allow the configurations to relax away from overlap in two ways.  One on hand, the chains trajectories can tip radially away from the high density region, as is described by a $\beta=1$ configuration.  We could further allow the brush to compress by imposing a sinusoidal variation of the brush height,  $H_b (u) = H_b (1+ \nu \sin u)$.  For configurations without overlap, we set $\nu = 0$, and otherwise, we choose the smallest $\nu$ such that the brush does not overlap.  Indeed, we find over a large range of parameters the free energy favors uncompressed states, with $\nu =0$.
By suitably rescaling $F_{b,st}$, the dimensionless corona stretching per chain, $\mathcal{B}[\cH_b(u), \ep]$, can be defined as
\begin{equation}
\label{eq: bst}
\frac{F_{b,st}}{n k_ T} = \frac{3 R_c^2}{N_b a^2 \pi} \frac{\mathcal{B}[\cH_b,\epsilon,\beta]}{\cV(\epsilon)}, 
\end{equation}
where $\cH_b=H_b/R_c$ is the height of the brush measured in units of the core size.

Finally, we assume that each chain has a fraction $\alpha$ of charged monomers equally distributed along the chain.  That is, we assume that the charge distribution is quenched.  The counterion contribution to the free energy is then ~\cite{jusufi}
\begin{eqnarray}
\frac{F_{ci}}{k_B T} &=& \rho_0 \int_{V_b} dV~\Big[ c_{+} \big (\ln \left(c_{+} \rho_0\right) -1 \big) \\
& & + c_{-} \big( \ln \left(c_{-} \rho_0\right) - 1\big)  - 2 c_s\big( \ln (c_s\rho_0) -1  \big) \Big],\nonumber
\end{eqnarray}
where $c_{\pm}$ is the position-dependent concentration of positive and negative ions, $c_s$ is the concentration of ions in bulk solution, and the integral is over the volume of the brush.  The mean-field counterion densities should, properly, come from solving the Poisson-Boltzmann equation for the charge distribution described by  $\alpha c_m(\textbf{r})$, where $c_m({\bf r})$ is the local concentration of monomer in the brush which depends on the brush geometry (see Appendix).  To derive this expression, we have assumed local electroneutrality, $c_{+} - c{-} = \alpha c_m$~\cite{borisov-zhulina}, which is valid in the osmotic regime~\cite{jusufi}.  From the Poisson-Boltzmann equation, we also have the relation $c_{+} c_{-} = c_s^2$ for a $1:1$ electrolyte.  This allows us to derive expressions for the ion concentrations in terms of $c_m$ and $c_s$:
\begin{equation}
\label{eq: charge}
c_\pm({\bf r}) = \pm \frac{\al c_m({\bf r})}{2} + \frac{1}{2}\sqrt{\big( \al c_m({\bf r})\big)^2 +  c_s^2}.
\end{equation}
Strictly speaking, we are considering the limit of the Poisson-Boltmann theory in the limit that the Debye screening length, $\kappa^{-1}$, is small compared to brush dimensions \cite{fyl}.  This is valid for the case when the concentration of free charge (counterions and salt) in the brush is sufficiently large.  We note that the expression for the osmotic pressure used in a similar mean-field calculation for polyelectrolyte copolymers by Borisov and Zhulina \cite{borisov_zhulina_diblock} can be obtained by taking the $c_s \gg c_m$ limit of the above expressions.

In general, one expects that some fraction of the counterions will be condensed onto the chains -- we can absorb this into a redefinition of the fraction of charged monomers $\alpha$ and assume that $\alpha$ does not change overly much when the interface geometry changes.  Additionally, some fraction of the counterions will escape the corona, an effect we also ignore in our approximations.  This will lead to region near the edge of the brush which is non-neutral.  However, we expect that the size of such a region will be proportional to the electrostatic screening length, which is small compared to the brush size.  It seems doubtful that these approximations will change the overall qualitative conclusions of this paper, though they are likely to play a quantitative role in some aspects of our calculation.  This will be discussed where appropriate in the following.

Given the energetics described here we obtain mean-field results by minimizing the free energy per chain over unconstrained parameters.  For the case of the equilibrium calculations discussed below, we minimize all free parameters.  The counterion free energy density does not depend on the core size $R_c$, allowing us to minimize over $R_c$ analytically, only on the ratio of brush height to core size, $\cH_b$.  We retain three parameters to be minimized over numerically: the brush height $h$, the undulation parameter $\epsilon$, and the brush chain direction $\beta$.  This gives us
\begin{equation}
R_c^3 = \frac{\gamma \cA(\epsilon) N_c^2 a^2}{\rho_0 \{ \mathcal{S}(\epsilon) \pi/8 + 3 f \mathcal{B}(\cH_b,\epsilon,\beta)/[ \pi (1-f)] \}},
\end{equation}
where $f = N_c/(N_c+N_b)$.

\begin{figure}
\begin{center}
\resizebox{3.25in}{!}{\includegraphics{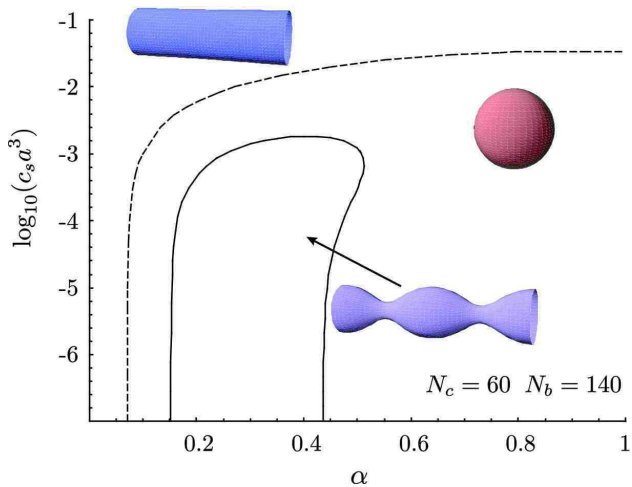}}
\caption{The phase behavior for hairy micelles ($f=0.3$).  The free energy per chain of cylinders and individual spheres is equal along the dashed line.  The sphere phase is stable in the lower right portion of the phase space, that is, at high $\al$ and low $c_s$.  Deep in the regime in which spheres have the lowest free energy, the cylinder state becomes unstable to the formation of undulated cylinders; this region is surrounded by the solid line.}
\label{fig:phasediagramf0p3}
\end{center}
\end{figure}

\begin{figure}
\begin{center}
\resizebox{3.25in}{!}{\includegraphics{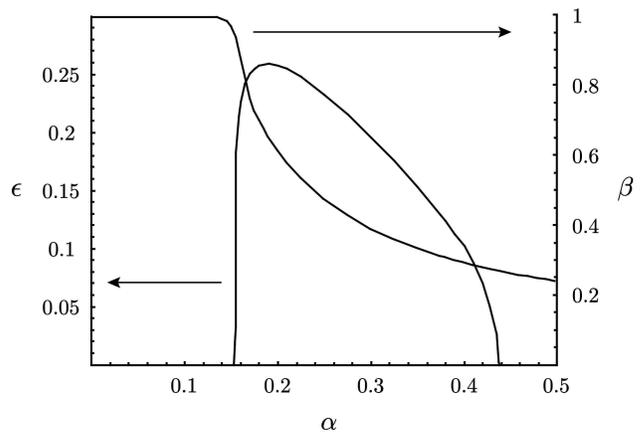}}
\caption{Plot of $\epsilon$ and $\beta$ as a function of $\alpha$ for hairy micelles ($f=0.3$) at salt concentration $c_s a^3 = 10^{-4}$.  As $\alpha$ increases $\beta$ is decreasing, indicating that the chains are tipping up radially.}
\label{fig:epvsalpha}
\end{center}
\end{figure}

\section{Results}
\label{sec:results}
\subsection{Metastability of Undulated Cylinders}
\label{sec:UC}
We computed mean-field phase diagrams for ``crew-cut" ($f=0.3$), symmetric ($f=0.5$) and ``hairy" ($f=0.7$) micelles.  The free-energy minimization was performed for fixed charge fraction, $\alpha$, and salt concentration, $c_s$.  We assume for simplicity that $\rho_0=a^{-3}$, and we consider copolymers composed of $N = N_c+N_b = 200$ monomers per chain, roughly equivalent to the experiments of Bendejacq \textit{et al}~\cite{bendejacq}.  We set the surface tension to $\gamma a^2= 5 k_B T$ and find that hairy micelles ($f=0.3$) are computed to have core radii in the range of 5 to 10 nm (assuming $a=$ 5 \AA), which is in good agreement with the experimental observations of ref. \cite{bendejacq}.

The phase behavior for hairy micelles ($f=0.3$) as a function of charge fraction $\alpha$ and dimensionless salt concentration $c_s a^3$ is shown in Fig.~\ref{fig:phasediagramf0p3}.  The free energy per chain of cylindrical and spherical micelles are equal along the dashed line.  At high salt and low polymer charge cylindrical micelles are stable, while at low salt and high charge spherical micelles become stable.  However, cylindrical micelles that persist into the stability region of spheres become unstable to undulation.  These undulations are driven by the frustration induced by the counterions, which prefer interfaces with positive Gaussian curvature in order to increase their free volume.  However, this tendency toward the formation of spheres must be balanced by the significant increase in free ion concentration in the regions of negative Gaussian curvature which is only partially mitigated by the migration of chains toward the positive curvature region (i.e. surface density of chains is reduced in the saddle regions; see Fig.~\ref{fig:model} for a schematic description).  Therefore we conclude that the undulated cylinder shape is \textit{metastable}.

\begin{figure}
\begin{center}
\resizebox{3.25in}{!}{\includegraphics{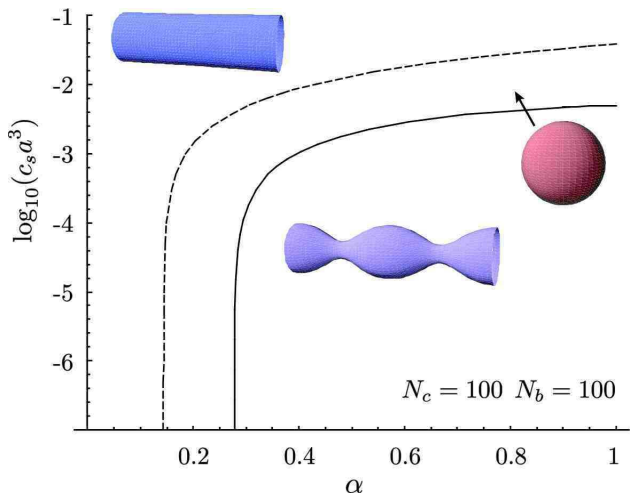}}
\caption{The phase behavior for symmetric micelles ($f=0.5$).  The free energy per chain of cylinders and individual spheres is equal along the dashed line.  Here, the unduloid state is metastable below and to the right of the dark line.}
\label{fig:symphase}
\end{center}
\end{figure}

In Figure \ref{fig:epvsalpha} we plot $\ep$ and $\beta$ as function of $\al$ for fixed salt concentration ($c_s a^3=10^{-4}$). When the cylinders first become unstable at low charge fraction, the undulated cylinders have eccentricities on the order of $\epsilon \approx 0.3$ and $\beta \approx 1$ so that the chains are almost nearly directed normal to the interface.  As the charge fraction is increased, the increased osmotic pressure within the brush increases its height (see Fig. \ref{fig:epvsalpha}).  This effect leads to an increase in the monomer and counterion concentration in the negative curvature regions.  The subsequent rise in osmotic pressure leads to a tendency of the chain trajectories to tip toward the radial direction.  Once $\beta$ decreases to zero and the brush chains extend radially from the cylinder axis, the counterions no longer gain any entropy from the extra interfacial curvature of the unduloid.  In fact, the core free energies, the core stretching and interfacial energy, are minimal for the $\ep=0$ cylindrical state.  And since the brush free energy is not lower for an unduloid state with $\beta=0$ than the $\ep=0$ state, $\ep$ decreases back to 0 for large $\al$.  The boundary of the undulated cylinder region at high charge fraction is determined by the condition $\epsilon=0$ (see Fig.~\ref{fig:epvsalpha}).

\begin{figure}
\begin{center}
\resizebox{3.25in}{!}{\includegraphics{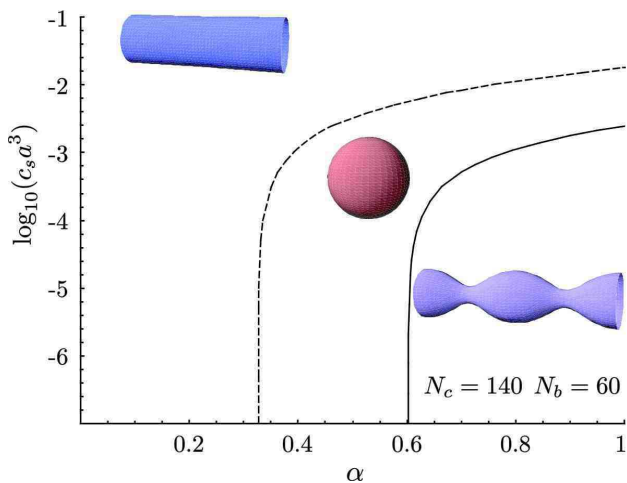}}
\caption{The phase behavior for crew-cut micelles ($f=0.7$).  The free energy per chain of cylinders and individual spheres is equal along the dashed line.  Here, the unduloid state is metastable below and to the right of the dark line.}
\label{fig:crewphase}
\end{center}
\end{figure}

The phase behavior of symmetric ($f=0.5$) and crew-cut ($f=0.7$) micelles is shown in Figs.~\ref{fig:symphase} and~\ref{fig:crewphase} respectively, and is very similar to the hairy case.  As we progress toward the crew-cut limit, the CS transition moves toward larger charge fractions and lower salt concentrations.  This may be due to the core energy which prefers cylindrical micelles over spheres and is dominated by the interfactial term, (\ref{eq: int}), becomes more important as the size of the core block, $N_c$, increases.    As the core chain length increases, the undulated cylinder behavior no longer disappears at large charge fraction for $\al < 1$.  As in the case of hairy micelles, however, spheres become energetically favorable well before the UC region in both cases.

\subsection{End caps}
\label{sec:endcaps}
\begin{figure}
\begin{center}
\resizebox{3.25in}{!}{\includegraphics{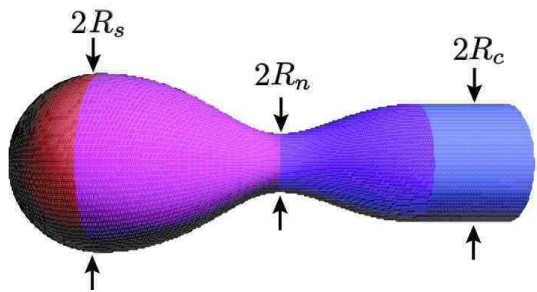}}
\caption{The end of a cylinder is a piecewise continuous concatenation of regions with constant interfacial curvature.  A hemispherical cap is attached to a pair of unduloids which are attached to a cylinder.}
\label{fig:endansatz}
\end{center}
\end{figure}

Given our model for the free energy of micelles constructed from the unduloid surfaces, we build an {\it ansatz} for the shapes of the ends to micelles of cylindrical micelles from consider piecewise constant mean curvature surfaces.  This is done by ``gluing'' two unduloids quarter-periods and a spherical hemisphere to the end of a cylinder with radius $R_c$ (see Fig.~\ref{fig:endansatz}).  One of the unduloid pieces changes radius from $R_c$ to $R_n$, the neck radius.  The second piece increases the radius from $R_n$ to the spherical cap radius, $R_s$.  Since we assume that the micelles are effectively infinite, the end pieces can increase in volume by pulling chains from the infinite cylindrical portion, and thus, raise or lower the energy of that particular chain relative to its energy in the cylinder.  Thus, to compute the free energy of these ends we consider that the ends are in contact with an infinite source of polymer chains at fixed chemical potential, which is set by the free energy per chain of the infinite cylindrical state.  The cylinder radius, $R_c$, is set to the equilibrium value of an infinite cylinder, and we minimize the free energy of the end over all remaining unconstrained parameters.

\begin{figure}
\begin{center}
\resizebox{3.25in}{!}{\includegraphics{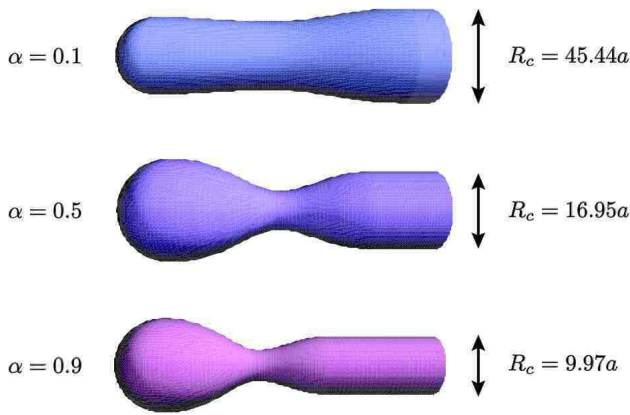}}
\caption{Typical end shapes for various charge fractions, in particular, for hairy micelles ($f=0.5$) at salt concentration, $c_s a^3=10^{-4}$.  The cylinder radius $R_c$ is set by the equilibrium cylinder radius.}
\label{fig:endshape}
\end{center}
\end{figure}

Many of the features of the characteristic end shapes can be explained by considering the free energy per chain in different geometries.  In the region of cylinder stability (at low charge and high salt), chains in the endcap have a higher free energy than those in the bulk cylinder, and therefore preferentially migrate into the cylinder region.  This has the effect of decreasing the radius of the spherical endcap, $R_s$, giving the appearance of the interface at $\al=0.1$ in Fig.~\ref{fig:endshape}.  In the spherical stability regime (at high charge and low salt), we find that the endcap, $R_s$, is generically larger than the cylinder radius, $R_c$.  Since in this regime the free energy of chains within the spherical bulges is lower than for chains with the cylindrical portion, chains preferentially migrate toward the spherical cap, increasing the radius of end cap, $R_s$.  To best approximate the preferred spherical geometry, the neck radius $R_n$, pinches down to well below $R_c$.  This procedure yields a spherical cap radius $R_s$ very close to the ideal spherical micelle size.  This characteristic ``bulbous"  end shape has been widely observed in a number of polymeric micelle systems \cite{eisenberg2, jain_bates_macro_04} , and is similar to the ends conjectured to appear in the context of bicontinuous phases of emulsions~\cite{hypercruiser}.  However, despite the preference for a spherical micelle configuration indicated by these bulbous end shapes, the free energy always has a minimum at finite neck size.  Indeed, the high osmotic pressure induced by the large ion concentration in the saddle regions of the charged brush domain lead to sizable energy barriers to ``pinching off'' ($R_n \rightarrow 0$).  These barriers are discussed in detail below.

\subsection{Energy Barriers to CS Transition}
Since undulated cylinders are a metastable state, we consider here the free energy barriers governing the eventual breakup of the micelles into spheres.  Diblock copolymer kinetics is a subject of ongoing theoretical~\cite{halperin_alexander_macro_89,  dormidontova_macro_98} and experimental~\cite{esselink_macro_98, lund_willner_prl_06} research.  Since diblocks in solvent are the polymeric equivalent of amphiphilic surfactants, it is instructive to compare the kinetics of their respective micellar distributions.  In surfactants, there is a substantial fraction of free molecules and the primary kinetic mechanism for morphological changes is the Aniansson-Wall (AW) mechanism~\cite{AWdynamics1, AWdynamics2}.  Briefly, the exchange of single surfactant molecules (unimers) from micelle to bulk solution is fast enough to allow the direct formation of spheres and the subsequent depletion of molecules from cylindrical to spherical micelles.  In the higher molecular weight copolymers, however, the concentration of free chains is significantly suppressed \cite{jain_bates_macro_04} and, consequently, the AW unimer exchange mechanism is drastically slowed \cite{lund_willner_prl_06}.  Instead, Dormidontova has found that fission-fusion events of micelles likely play a significant role in the relaxation kinetics of copolymers solutions~\cite{dormidontova_macro_98}.  This suggests that the kinetic pathway connecting cylindrical and spherical micelle distributions may be dominated by states of fixed polymer number and intermediate interfacial geometry, such as undulated cylinders \cite{jain_bates_macro_04}.

\begin{figure}
\begin{center}
\resizebox{3.25in}{!}{\includegraphics{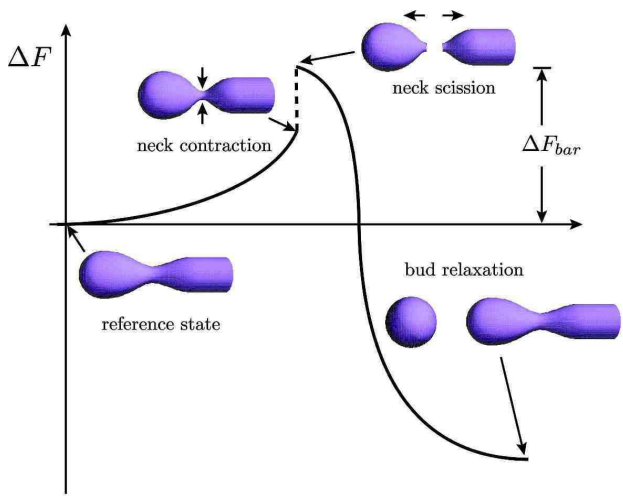}}
\caption{A skematic plot of the free energy along the assumed kinetic pathway for cylindrical ends pinching off to for spherical micelles.  The horizontal axis is some generalized reaction coordinate which carries the micelle end through the CS transition along the lowest energy pathway.  The transition occurs in three stages:  neck contraction below the equilibrium size, $R_n$, neck scission, and bud and and end relaxation.  Presumably, the scission must expose the insoluble blocks to solvent, and therefore it is visualized as a discontinuous jump in the free energy of the end.}
\label{fig:budpath}
\end{center}
\end{figure}

We turn our attention to the calculation of the free energy barriers for the fission of a cylindrical (or undulated) micelle end into spherical micelles.  Our end shape {\it ansatz} allows us to compute the energetic barrier for budding a sphere from the end of a cylindrical micelle by continuously tuning the neck size, $R_n$.  A schematic free energy plot for this pathway is depicted in Fig.~\ref{fig:budpath}.  We consider a pathway which is divided into three stages:  (1) contraction of the neck radius, (2) scission of the neck, and (3) bud relaxation.  While, in principle, we could imagine a pathway in which $R_n$ contracts continuously to zero (or at least some molecular size), we find that allowing for neck scission lowers the over barrier considerably.  Therefore, once the neck size has decreased below some critical size, the bud neck breaks, thereby exposing an area of the core $2 \pi R_n^2$ to solution.  At the point of neck scission, we estimate the cost of exposing core monomers to solution as $\gamma$, the interfacial surface tension, times the exposed surface area.  To compute the minimum energy pathway within our {\it ansatz} we minimize the free energy barrier, $\Delta F_{bar}$, over this critical neck radius.  

\begin{figure}
\begin{center}
\resizebox{3.25in}{!}{\includegraphics{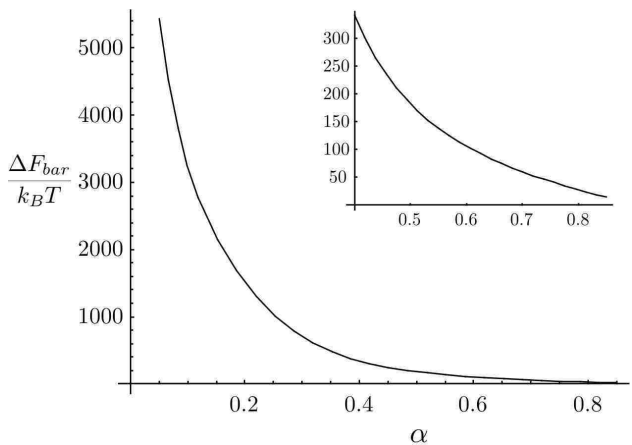}}
\caption{A plot of the CS barrier height as a function of $\alpha$ for hairy micelles ($f=0.3$) at low salt, $c_s a^3=10^{-6}$.  The inset displays the small barrier height region at large charge fraction in rescaled coordinates for convenience.}
\label{fig:barrier}
\end{center}
\end{figure}

Our calculation for the barrier height for end fission of spherical micelle from a cylinder end is plotted in figure \ref{fig:barrier} for hairy micelles ($f=0.3$) and for low salt ($c_s a^3=10^{-6}$).  We find that $\Delta F_{bar}$ {\it decreases} for increased $\alpha$.  When $\alpha$ is small, the barrier heights can be quite large (thousands of $k_B T$).  For larger $\alpha$, however, the barrier heights fall precipitously to tens or hundreds of $k_B T$.  In comparison to the case neutral copolymer melts~\cite{matsen_jcp_01}, the free energy barriers computed here are quite large, up to 1000 times larger.

A simple consideration of the free energy scale set by the brush free energy shows that the excluded volume interaction induced by the brush electrostatics should indeed lead to free energy barriers of this magnitude.
While excluded volume effects are screened in molten polymer brushes, the CS transition is frustrated by the increased osmotic pressure in the neck region of the end bud.  Typically, such an end bud will contain $10^2-10^3$ polymer chains, and each of those chains entrains $\al N_b$ counter ions within the dissolved brush.  Given that each of those counterions represents roughly $k_B T$ worth of free energy due to the entropy lost in confinement, it is then not surprising that such a topological transition could cost thousands of $k_B T$.   Nonetheless, we see from Figure \ref{fig:barrier} that changing the effective charge of the polyelectrolyte block can {\it decrease} the barrier at least two orders of magnitude.  The qualitative explanation of this effect is simple. As $\al$ increases, the equilibrium size of the micelles decreases, indicating that fewer chains comprise the micelle ends, leading to an overall decrease in $\Delta F_{bar}$.

\begin{figure}
\begin{center}
\resizebox{3.25in}{!}{\includegraphics{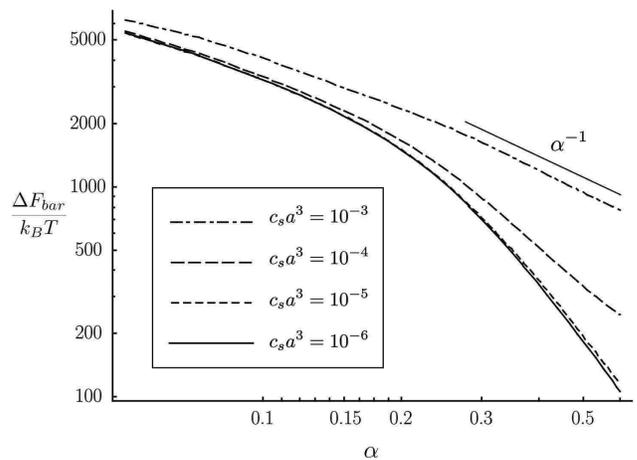}}
\caption{The CS barrier height for hairy micelles ($f=0.3$) as a function of $\alpha$ on a log-log scale for a number of different salt concentrations.}
\label{fig:logbarrier}
\end{center}
\end{figure}

The barrier heights for hairy micelles are plotted on a log-log scale in figure~\ref{fig:logbarrier} for a number of different salt concentrations.  We find two characteristic regimes of power law behavior, one at low charging and one at high charging.  At high salt concentration and large $\alpha$, we find that the barrier height decreases as $\alpha^{-1}$.  In this regime, the decrease in neck radius $R_n$ is small and the free energy barrier is dominated by direct scission.  Here, the dependence on $\alpha$ comes directly from the equlilibrium scaling of the cylinder radius with $\al$ at high salt in the limit $H_b\gg R_c$, $R_c \sim \alpha^{-1/2}$ \cite{netz}.  Largely, we expect the equilibrium neck radius to scale with $R_c$,  Therefore, direct scission predicts a free energy barrier, $\Delta F_{bar} \sim \gamma  R_n^2$, which yields the observed power law.

\section{Discussion}
\label{sec:discussion}
We have developed a free energy expression for diblock copolymer cylinders with a hydrophobic block and a charged corona, and used this to understand the stability of various axisymmetric structures.  We use a family of constant mean curvature surfaces called unduloids to parametrize possible interfaces from cylinder to a necklace of spheres.  This {\it ansatz} allows us to probe the full, non-linear dependence of the free energy on the micelle surface geometry.  Our results indicate that there is a metastable undulated cylinder phase, as observed in experiments~\cite{bendejacq} which occurs generically in parameter ranges where the free energy per chain for spherical micelles is lower still.  We have studied the shape of the spherical endcaps and found, in particular, that in the regime in which cylinders are stable, the ends show a distinct decrease in radius below the equilibrium value.  When spherical micelles are preferred in equilibrium, the spherical cap radius is larger than the cylinder radius -- and is close to the preferred spherical micelle size.  We use these results to compute dependence the free energy barriers for budding spherical micelles from cylindrical micelles on polymer charge, $\al$, and solution salt concentration, $c_s$.

Our calculation of the free energy is variational and therefore our values for the free energy barrier to budding, $\Delta F_{bar}$, necessarily provide upper bounds.  However, certain conclusions from this study are robust.  Indeed, these results along with the fact that the AW step-wise kinetic mechanism is suppressed in polymeric systems \cite{dormidontova_macro_98} strongly support the suggestion that distributions of polymer micelles in selective solvent ``non-ergodic," with micelles essentially trapped within a fixed, quenched topology \cite{jain_bates_macro_04}.  In contrast with neutral copolymer solutions, however, we find here a strong, tunable dependence of barrier height on polymer charge and salt concentration.  Kramers kinetic theory predicts that the CS transition rate, $k_{{\rm CS}}$ should depend on the the free energy as $k_{{\rm CS}} \propto e^{-\Delta F_{bar}/k_B T}$ \cite{kramers}.
At small $\alpha$, the CS pathway investigated here is kinetically arrested by large barrier heights.  Supposing the AW unimer exchange mechanism is also sufficiently slow, these large barrier heights suggest that the metastable undulated cylinder regime may be observed.  Such undulated cylinders have been observed experimentally in a number of amphiphilic polymeric systems, both charged and uncharged~\cite{bendejacq, eisenberg1, jain_bates_macro_04}.  At large $\al$ and low $c_s$, $\Delta F_{bar}$ may fall to sufficiently low values that the transition to spherical micelles may proceed on experimental time scales.  This suggests that the observed ``phase boundary'' between cylindrical and spherical micelles in ref.~\cite{bendejacq}, may reflect that fact that these kinetic barriers depend quite strongly on $\al$ and $c_s$.  Given that $\Delta F_{bar}$ varies by two orders of magnitude, we expect an extremely sharp dependence of the $k_{{\rm CS}}$ on $\al$ and $c_s$ over the same parameter range.

The geometry of self-assembled structures is often cast in terms of constant mean curvature (CMC) surfaces~\cite{CMCdiblock,hypercruiser}.  More generally, the energy of any interface can be expanded in powers of the curvature, resulting in the Helfrich energy,
\begin{equation}
\label{eq:Helfrich}
E = \gamma A_{tot} + \frac{\kappa}{2} \int dA~\left(H-H_0 \right)^2+\bar{\kappa} \int dA~K,
\end{equation}
where $A_{tot}$ is the total micelle interfacial area, $H=(1/2) (1/r_1 + 1/r_2)$ is the mean curvature, $K = 1/(r_1 r_2)$ is the Gaussian curvature, $H_0$ is the spontaneous curvature, which comes from the molecular asymmetry of the chains and the corona-solvent interaction.  The parameter $\gamma$ is the surface tension.  Finally, $\kappa$ and $\bar{\kappa}$ are effective bending rigidities for the mean and Gaussian curvatures respectively, which must be derived from a microscopic model.  For $\kappa=\bar{\kappa} = 0$, this energy describes the shape of soap bubbles, and CMC surfaces minimize the energy in this limit.  In the context of diblock copolymers~\cite{nonCMCdiblocks1, nonCMCdiblocks2} or fluid membrane shapes \cite{toan_ajay}, the additional curvature terms modify the shape and stability of various structures.  Higher order curvature terms are typically neglected as being subleading, but we will show equation (\ref{eq:Helfrich}) fails to describe our results, even qualitatively, without them.

We restrict ourselves to interfaces belonging to the unduloid family of surfaces capped at both ends.  The energy reduces to
\begin{equation}
E = \left[\gamma + \frac{\kappa}{2} \left(H-H_0\right)^2\right] A_{tot}(\epsilon) - 4 \pi \bar{\kappa},
\end{equation}
where we note that a cylinder with two ends has $\int dA~K = - 4 \pi$ from the Gauss-Bonnet theorem. If we choose the length of the cylinder to conserve the volume, the total area is given by $A_{tot} (\epsilon) = [\cA(\epsilon)/\cV(\epsilon)] V_{tot}$, where $\cA(\ep)$ and $\cV(\ep)$ are the reduced area and volume of a half undulation wavelength (see appendix) and $V_{tot}$ is the volume of the cylinder.  The only dependence on $\epsilon$ comes from the volume and area of the interface (recall that $H$ does not change with $\epsilon$) which, at fixed $R_c$, is minimum for $\epsilon = 1$ no matter the value of spontaneous curvature.  This reflects the fact that spheres minimize the area for a fixed volume, and indeed, this is a manifestation of the well-known Rayleigh instability of fluid interfaces \cite{safran}.  The spontaneous curvature only sets the overall size scale of the micelles, not their topology.

We conclude that higher order curvature terms to equation (\ref{eq:Helfrich}) are required to understand the local stability of undulated cylinders in terms of universal geometrical measures of the block interface~\cite{hypercruiser, fournier}.  This is, perhaps, not surprising since higher order corrections introduce length scales on the order of $R_c$ for axisymmetric shapes~\cite{fournier}.  In the case of vesicles, where equation (\ref{eq:Helfrich}) is known to apply, these corrections are controlled by the membrane thickness~\cite{fournier, safran}.  We leave the development of a geometrical shape equation for diblock copolymer micelles to future work.

\begin{acknowledgements}
The authors are pleased to acknowledge discussions with D. Bendejacq, V. Ponsinet, P. Pincus, and R. Kamien.  GMG was supported by NSF Grant DMR04-04507. CDS was supported by NSF Grants DMR01-29804, DMR05-47230, and the Pennsylvania Nanotechnology Institute.  In addition, both authors were supported under NSF Grant DMR05-20020 (MRSEC), ACS Petroleum Research Fund, and a generous gift from L.J. Bernstein.
\end{acknowledgements}
\appendix

\section{Unduloid Geometry}
In this Appendix we present details concerning the geometry of unduloid surfaces which will be useful in our free energy calculation.  Given the parametric representation of ${\bf X}(u, \phi)$ by eqs. (\ref{eq: z}) and (\ref{eq: r}) we have the differentials,
\begin{equation}
dr = \frac{R_c ( \ep \cos u) du}{\sqrt{1+\ep^2+2 \ep \sin u}} \ ,
\end{equation}
and 
\begin{equation}
dz = \frac{R_c(1+\ep \sin u) du }{\sqrt{1+\ep^2+2 \ep \sin u}} \ .
\end{equation}
The tangent vectors for the two principle directions of the surface are given by,
\begin{equation}
d {\bf X}_\phi = r(u) d \phi \hat{\phi}
\end{equation}
and 
\begin{equation}
d {\bf X}_u = \frac{R_c du}{\sqrt{1+\epsilon^2+2 \epsilon \sin u} } \big[(1+\epsilon \sin u) \hat{z}+\epsilon \cos u \hat{r} \big] \ .
\end{equation}
From these we compute the differential element of area,
\begin{equation}
dA=|d {\bf X}_\phi \times d {\bf X}_u|=R_c^2\sqrt{1+\ep^2+2 \ep \sin u} du d \phi \ ,
\end{equation}
and the surface normal,
\begin{equation}
\hat{{\bf n}}=\frac{d {\bf X}_\phi \times d {\bf X}_u}{|d {\bf X}_\phi \times d {\bf X}_u|}=\frac{\big[-\epsilon \cos u \hat{z} + (1+\epsilon \sin u) \hat{r} \big]}{\sqrt{1+\epsilon^2+2 \epsilon \sin u} } \ .
\end{equation}
Since the surface is periodic we concern ourselves only with a half period of the undulation, $u \in [-\pi/2,\pi/3]$.  The volume per half period can be computed,
\begin{eqnarray}
\nonumber
V(\ep)&=&\pi \int_{-\pi/2}^{\pi/2} du \frac{dz}{du} r^2(u) \\
&=& \pi R_c^3 \int_{-\pi/2}^{\pi/2} du \sqrt{1+\ep^2 + 2 \ep \sin u}(1+\ep \sin u) \  . \nonumber  \\ .
\end{eqnarray}
The dimensionless volume, $\cV(\ep)$, is then defined by $V(\ep) \equiv \pi^2 R_c^3 \cV(\ep)$.  Finally, we surface area of a half period is computed by,
\begin{equation}
A(\ep)=2\pi R_c^2 \int_{-\pi/2}^{\pi/2} du \sqrt{1+\ep^2 + 2 \ep \sin u} \ ,
\end{equation}
and the dimensionless area, $\cA(\ep)$, is defined by $A(\ep) \equiv 2 \pi^2 R_c^2 \cA(\ep)$.

\begin{figure}
\begin{center}
\resizebox{2.75in}{!}{\includegraphics{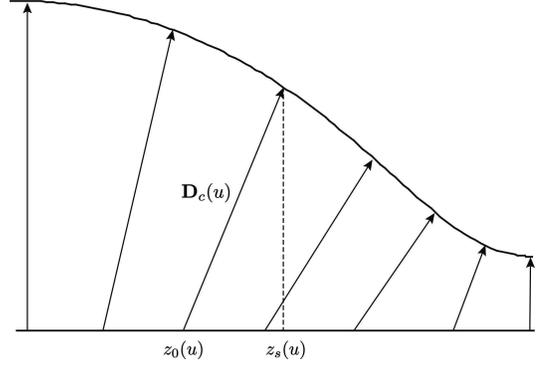}}
\caption{A schematic representation of the normal trajectory {\it ansatz}.}
\label{fig: map}
\end{center}
\end{figure}

\section{Micelle Core Energetics}
In this section we compute the core stretching (\ref{eq: cst}) .  As described in Sec. \ref{sec:model}, we divide the core region into infinitestimal wedges, which contain polymer chains extending in the same direction.   We assume that the core chains (and wedges) extend along the surface normal.   As shown in Figure \ref{fig: map} the vector, ${\bf D}_c(u)= r(u) \hat{r}- \frac{dr}{dz} r(u) \hat{z}$, describes the extension of the wedge. From the normal stretching condition, we have the mapping from a point on the surface to a point along the $z$ axis, located a $z_0(z)$ (see Figure \ref{fig: map}).  This mapping is given by,
\begin{equation}
z_0(u) = z(u)+\frac{d r}{dz} r(u) \ .
\end{equation}
In general, this mapping must be consistent with the constraint the chain paths (and wedges) do not cross.  This means that the mapping must be one-to-one.  This constraint is satisfied for $\frac{dz_0}{dz} \geq 0$.  This yields an inequality which must be satisfied by surfaces we consider,
\begin{equation}
-\bigg[\frac{d^2 r}{d z^2}r+\Big(\frac{dr}{dz}\Big)^2\bigg] \leq 1\  .
\end{equation}
It is not difficult to show that all unduloid surfaces satisfy this inequality.

\begin{figure}
\begin{center}
\resizebox{2.75in}{!}{\includegraphics{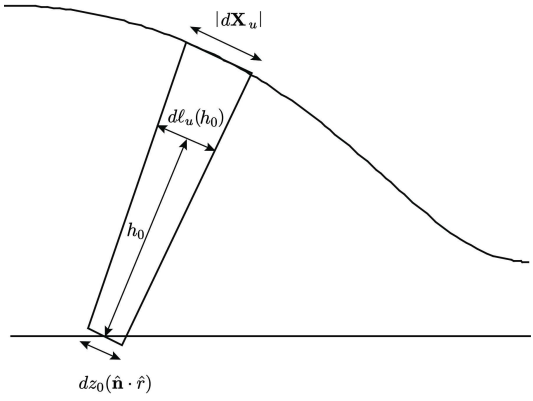}}
\caption{A differential wedge extending along the mean chain trajectory.}
\label{fig: core}
\end{center}
\end{figure}

\begin{figure}
\begin{center}
\resizebox{3in}{!}{\includegraphics{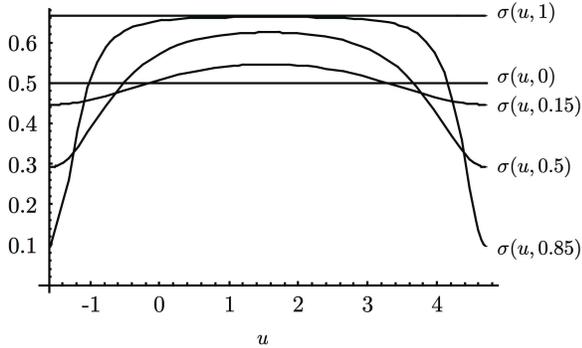}}
\caption{Plots of the surface chain density for various unduloid shapes.  The density is plotted in units of $\rho_0 R_c / N_c$.}
\label{fig: sigma}
\end{center}
\end{figure}

To compute the stretching integral defined in eq. (\ref{eq: cst}), we must compute the volume distribution within the wedge, which we do by considering the dimensions of the wedge in the $d {\bf X}_u$ and ${d {\bf X}_\phi}$directions as a function of distance to the $z$ axis, $h_0$.  We will denote these directions by $d \ell_u (h, u)$ and $ d \ell_\phi (h,u)$, respectively (see Figure \ref{fig: core}).  These are given by,
\begin{equation}
d \ell_u(h_0,u) = dz_0 (\hat{{\bf n}} \cdot \hat{r} )+ \frac{h_0}{D_c(u)}\big(R_c du - dz_0 (\hat{{\bf n}} \cdot \hat{r} ) \big) \ ,
\end{equation}
and 
\begin{equation}
d \ell_\phi (h_c,u) =  \frac{h_0}{D_c(u)} r(u) d\phi \ ,
\end{equation}
where $D_c(u)=|{\bf D}_c(u)|$.  From this we compute the area distribution in the core wedges by $dA_c(h_0,u)= d\ell_u (h_0, u) d \ell_\phi (h_0,u)$.  
With this the volume per core wedge can be computed,
\begin{equation}
dv_c(u)= \int_0^{D_c(u)} dh_0dA_c(h_0,u) \
\end{equation}
and the surface density, $\sigma(u)$, of chains reaching the interface at, 
\begin{equation}
\sigma(u)= \frac{dv_c(u) \rho_0}{N_c dA(u)} .
\end{equation}
We plot $\sigma(u)$ for various unduloid shapes in Figure \ref{fig: sigma}.  For cylinders ($\ep=0$) and spheres ($\ep=1$) the surface density is constant, while for intermediate shapes $\sigma(u)$ drops in the saddle regions.

Finally we compute the stretching free energy for a half period of the core region,
\begin{eqnarray}
\nonumber
\frac{F_{c,st}(\epsilon,R_c)}{k_B T} &=& \frac{3 \pi^2 \rho_0}{8 N_c^2 a^2} \int_0^{2 \pi} \!\! d \phi \int_{-\pi/2}^{\pi/2} \!\!  du \int_0^{D_c(u)} \!\! dh_0 \\ 
\nonumber && \ \ \ \ \ \ \ \ \ \ \times (D_c(u)-h_0)^2 dA_c(h_0, u)  \\
& \equiv &  \frac{ \pi^3 \rho_0 R_c^5}{8 N_c^2 a^2} \mathcal{S}(\epsilon) \ ,
\end{eqnarray}
where the dimensionless stretching is defined on the last line.  Dividing this with the number of chains in a half period of core is $\pi^2 R_c^3 \cV(\ep) \rho_0/N_c$, we obtain eq. (\ref{eq: cst}).  

\section{Outer Brush Energetics}
\begin{figure}
\begin{center}
\resizebox{2.75in}{!}{\includegraphics{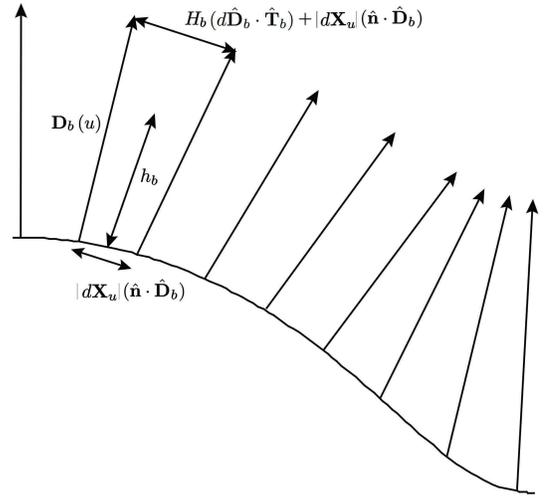}}
\caption{Geometry of coronal polymer brush.}
\label{fig:brushtrajectory}
\end{center}
\end{figure}

In Figure \ref{fig:brushtrajectory} we show the geometry of the chain trajectories in the outer polyelectrolyte brush.  The vector ${\bf D}_b(u)=H_b(u)\hat{{\bf D}}_b (\ep, \beta)$ gives the extension of brush chains from the interface surface at ${\bf X}(u, \phi)$ to the surface described by the tips of this outer brush at ${\bf X}(u, \phi)+ {\bf D}_b(u)$.  As described in Sec. \ref{sec:model} make the assumption that all chains are uniformly stretched, with their free ends constrained to tips of the brush.  Given this assumption, it is straightforward to compute the free energy cost of extending the brush within a half period of undulation,
\begin{eqnarray}
\nonumber
\frac{F_{b,st}}{k_B T} &=& \frac{3 \pi}{N_b^2 a^2}  du \int_{-\pi/2}^{\pi/2} \!\! du  H_b^2(u) \sigma(u) \frac{ d A}{du} \nonumber \\
& \equiv &  \frac{ 3 \pi \rho_0 R_c^5}{N_c N_b a^2} \mathcal{B}[\cH_b(u),\epsilon] \ .
\end{eqnarray}
Dividing by the number of chains in the core, we obtain eq. (\ref{eq: bst}).

Next we compute the brush monomer concentration, $c_m({ \bf r}$, which depends on the brush geometry.  Therefore, we need to determine the volume distribution as we did for the core brush.  In particular, we divide the outer brush into wedges extending along the mean polymer trajectory and we compute the area distribution within these wedges as we did in the case of the core (see Fig. \ref{fig:brushtrajectory}).  Here, we define $d \ell_\phi (u, h_b)$ as we did before, and $d \ell_u (h, h_b)$ is the dimension of the wedge associated with the direction perpendicular to $\hat{{\bf D}}_b$ which we denote  $\hat{ {\bf T}}_b=(\hat{{\bf D}}_b \cdot \hat{r} ) \hat{z}-(\hat{{\bf D}}_b \cdot \hat{z} ) \hat{r} $.  These can readily be computed from the geometry,
\begin{equation}
d \ell_u(h, h_b)= |d {\bf X}_u|(\hat{{\bf n}} \cdot \hat{{\bf D}}_b )+h_b ( d \hat{{\bf D}}_b \cdot \hat{{\bf T}}_b)\ ,
\end{equation}
and 
\begin{equation}
d \ell_\phi(u,h_b)=r(u) d \phi+h_b ( \hat{{\bf D}}_b \cdot \hat{r}) \ .
\end{equation}
We can compute the area available to polymers in the wedge at a distance, $h_b$, from the interface by $dA_b(u,h_b)= d \ell_u(u,h_b)d \ell_\phi(u, h_b)$.  If a given wedge leaves the surface from ${\bf X}(u, \phi)$, then it contains $\sigma(u) dA(u)$ polymer chains.  Given that we consider the corona chains to be evenly stretched throughout the brush, a chain leaves, $d h_b N_b/ H_b$ monomers between $h_b$ and $h_b+dh_b$, in a differential volume, $dA_b(u, h_b) dh_b$.  Therefore, the position dependent monomer concentration is,
\begin{equation}
c_m({\bf r})= \frac{ N_b \sigma(u) dA(u)}{H_b d A_b(u, h_b) }\ .
\end{equation}
We use this relation in eq. (\ref{eq: charge}) to compute the concentrations of free ions in the brush.

\end{document}